\documentclass{mn2e}
\usepackage{epsf,times}
\usepackage{amsmath}
\newcommand{\etal}{{et al}\/.}

\begin{document}
\title[Radio galaxies and cosmic rays]{Which radio galaxies can make
  the highest-energy cosmic rays?}
\author[M.J.~Hardcastle]{M.J.\ Hardcastle\thanks{E-mail: m.j.hardcastle@herts.ac.uk}\\School of Physics,
  Astronomy and Mathematics, University of
Hertfordshire, College Lane, Hatfield, Hertfordshire AL10 9AB}
\maketitle
\begin{abstract}
Numerous authors have suggested that the ultra-high energy cosmic rays
(UHECR) detected by the Pierre Auger Observatory and other cosmic-ray
telescopes may be accelerated in the nuclei, jets or lobes of radio
galaxies. Here I focus on stochastic acceleration in the lobes. I show
that the requirement that they accelerate protons to the highest
observed energies places constraints on the observable properties of
radio lobes that are satisfied by a relatively small number of objects
within the Greisen-Zat'sepin-Kuzmin (GZK) cutoff; if UHECR are protons
and are accelerated within radio lobes, their sources are probably
already known and catalogued radio galaxies. I show that lobe
acceleration also implies a (charge-dependent) upper energy limit on the
UHECR that can be produced in this way; if lobes are the dominant
accelerators in the local universe and if UHECR are predominantly
protons, we are unlikely to see cosmic rays much higher in energy than
those we have already observed. I comment on the viability of
the stochastic acceleration mechanism and the likely composition of
cosmic rays accelerated in this way, based on our current
understanding of the contents of the large-scale lobes of radio
galaxies, and finally discuss the implications of stochastic lobe
acceleration for the future of cosmic ray astronomy.
\end{abstract}

\section{Introduction}
\label{intro}

It has been known for many years (e.g. Hillas 1984) that the
large-scale structures of radio-loud active galaxies are possible
sites for the acceleration of the highest-energy cosmic rays yet to be
detected, the ultra-high-energy cosmic rays (UHECR) with energies
above a few $\times 10^{19}$ eV. Radio galaxy jets, hotspots and lobes
are particularly interesting to modellers, both because the
synchrotron emission by which we see them in the radio already implies
the presence of a high-energy particle population (albeit leptonic and
of much lower energies) and therefore of a particle acceleration
process, and because the physical conditions, in particular the
magnetic field strength $B$, can either be estimated from
equipartition or minimum energy arguments (Burbidge 1956) or, more
recently, determined directly from observations of inverse-Compton
emission (e.g. Hardcastle \etal\ 2002). It is thus reasonably easy to
say whether any given component of a radio galaxy is capable of
confining an energetic particle of a given energy and charge, a
necessary precondition in almost all models of particle acceleration.

The idea that radio-loud AGN might be the origin of the UHECR receives
some tentative support from, or is at least consistent with, recent
results from the Pierre Auger Observatory (PAO) suggesting that the
spatial arrival directions of UHECR above $6 \times 10^{19}$ eV are
correlated with local AGN (Abraham \etal\ 2007). The imposition of
this high low-energy cutoff on the cosmic rays ought to imply that
they have a relatively local (within $\sim 100$ Mpc) origin, since
UHECR at these energies coming from larger distances would suffer
strong attenuation due to interactions with the photons of the cosmic
microwave background radiation (the so-called Greisen-Zat'sepin-Kuzmin
or GZK cutoff; Greisen 1966) and also means that these UHECR undergo
the smallest possible deflection in the Galactic and intergalactic
magnetic fields. A particularly striking effect in the PAO data
released in 2007 was the spatial coincidence between several of the
UHECR and the position of the closest radio galaxy to us, Centaurus A
(e.g. Moskalenko \etal\ 2009). While it is not yet clear whether the
correlation with local AGN remains significant in the PAO data
collected since 2007, updated versions of the Abraham \etal\ (2007)
map appear to show a continued overdensity of UHECR around the
position of Cen A (e.g. Fargion 2009). Meanwhile, several authors have
suggested that the correlation between the arrival directions of UHECR
in the original PAO dataset and the positions of local radio-loud AGN
is at least as good as that with AGN in general (Nagar \& Matulich
2008; Hillas 2009).

How can specifically radio-loud AGN accelerate UHECR? It is of course
possible that they are accelerated on sub-parsec scales, comparable to
the scale of jet generation or initial collimation. The high photon
and magnetic field energy densities expected close to the active
nucleus provide important loss processes, but the acceleration
efficiency might also be higher. Many authors
have discussed mechanisms by which UHECR can be accelerated in the
nuclear regions of Cen A and of radio galaxies in general
(e.g.\ Kachelrie\ss, Ostapchenko \& Tom\`as 2009)
but these necessarily rely on assumptions about the physical
conditions close to the nucleus that are hard to test observationally.
In what follows I therefore focus on the larger-scale components of
radio-loud AGN.

{\it Direct} information about the leptonic particle acceleration
processes in radio galaxies, derived from observations in the optical
and X-ray where the synchrotron loss timescales are shorter than the
transport timescales from the nuclei so that {\it in situ} particle
acceleration is required, implies that particle acceleration must be
taking place in the hotspots of powerful double (Fanaroff \& Riley
1974 class II, hereafter FRII) radio galaxies, and in the kpc-scale
jets of the lower-power FRI class. FRII hotspots have traditionally
been modelled as the terminal shocks of the relativistic, internally
supersonic jet that extends up to Mpc scales in these objects (e.g.
Blandford \& Rees 1974; Heavens \& Meisenheimer 1987; Meisenheimer
\etal\ 1989), and, while optical and X-ray synchrotron evidence
complicates this picture (e.g. Prieto \etal\ 2002; Wilson, Young \&
Shopbell 2001; Hardcastle \etal\ 2007a) it seems clear that they are
particle acceleration sites. Moreover, their sizes and their magnetic
field strengths, which can be measured very well via the
inverse-Compton process in the most luminous systems where X-ray
synchrotron emission is not a contaminant (e.g. Harris \etal\ 1994;
Hardcastle \etal\ 2004) are certainly sufficient to allow UHECR to be
confined (Hillas 1984). However, the space density of FRIIs is very
low: we expect only a few within the GZK cutoff (for example, the
nearest FRII in the northern sky, 3C\,98, is at a distance of 134 Mpc)
and so their effect on the PAO sky above $6 \times 10^{19}$ eV is
negligible.

The numerically dominant population of radio galaxies, by several
orders of magnitude, within 100 Mpc is composed of low-power FRI
objects. Here the resolved particle acceleration region is typically
the 100-pc to kpc-scale inner jet. Several nearby FRI radio galaxies,
including Cen A (e.g., Hardcastle \etal\ 2003, 2007c; Goodger \etal\ 2010)
and M87 (e.g., Perlman \& Wilson 2005; Harris \etal\ 2006) have jets that are
comparatively strong sources of X-ray synchrotron emission, allowing
their particle acceleration properties to be studied in detail, while
the evidence is consistent with the idea that all powerful FRI jets
can accelerate leptons to the $>$ TeV energies required for X-ray
synchrotron emission (e.g. Worrall \etal\ 2001). The picture that
emerges from the X-ray observations is of a combination of strongly
localized particle acceleration, which may be due to small-scale
shocks, and a more diffuse process, which produces a
different X-ray spectrum (and therefore a different electron energy spectrum) and
which may therefore have different underlying acceleration physics. It
has been argued, most recently by Honda (2009), that the Cen A jet is
capable of accelerating protons to energies comparable to those of the
PAO UHECR, which of course implies acceleration of heavy nuclei to
even higher energies. This work relies on rather generous assumptions
about the sizes and magnetic field strengths of the acceleration
regions, though: as yet we have no direct constraint on the magnetic
field strength in FRI jets (although TeV inverse-Compton emission
should in principle provide one; Hardcastle \& Croston, in prep.).

This leaves us with the possibility of UHECR acceleration in the
lobes, the largest-scale components of both FRI and FRII radio
galaxies. At first sight these appear less promising candidates for
UHECR acceleration, since there is little direct evidence for {\it in
  situ} particle acceleration in the lobes. However, in the case of
the 600-kpc giant lobes of Cen A (Hardcastle \etal\ 2009, hereafter
H09) we showed that the high-frequency radio data from the {\it
  Wilkinson Microwave Anisotropy Probe} ({\it WMAP}) are consistent
with the idea that the lobes contain at least some relatively
energetic leptons; they do not rule out the idea that particle
acceleration is ongoing at some level. Similarly, while we do not as
yet have a robust inverse-Compton measurement of the magnetic field in
the lobes of any FRI radio galaxy, the available limits in the case of
Cen A constrain the field strength to be comparable to or greater than
the equipartition value. H09 argued that the known size, and the
limits on $B$, for the giant lobes meant that they could {\it confine}
protons of energies of order $10^{20}$ eV, and could therefore {\it
  accelerate} protons to such energies, provided that a relatively
efficient acceleration process was able to operate. We also showed
  that, provided that the energy index for the accelerated cosmic rays
  is relatively flat, the energetic requirements for the acceleration
  of the PAO UHECR plausibly associated with Cen A are trivially
  satisfied --- UHECR need only account for a small fraction of the
  total source energetics. Our preferred acceleration
mechanism involved scattering off relativistic turbulence within the
lobes, which requires the assumption that the internal energy density
is not dominated by thermal particles (see also
O'Sullivan \etal\ 2009) but is otherwise consistent with observations.
We will return to the question of particle content and lobe energetics
later in the paper, but in the next section I will show that a model
in which UHECR are accelerated in the giant lobes is unique in
providing some predictions for the spatial and energetic properties of
UHECR which may already be testable using the PAO data.

Finally, it should be noted that none of the above mechanisms are
mutually exclusive. In fact, it seems highly likely that, in a source
like Cen A, hadronic cosmic rays can be accelerated in the nucleus and
the kpc-scale jet as well as in the giant lobes. Particles accelerated
in the inner few kpc will eventually pass into the giant lobes and
will then be confined (and potentially accelerated) there for some
time before escaping. Hybrid models of this form potentially reduce
the problems of acceleration purely in the lobes, by providing a seed
population of cosmic rays at say $10^{17}$ -- $10^{18}$ eV and
therefore reducing the required UHECR acceleration time in the lobes.
A corollary of this, unfortunately, is that the ability of the giant
lobes to {\it confine} UHECR, irrespective of whether they can
accelerate them, implies that the UHECR will be {\it emitted} by a
source like Cen A on scales of the giant lobes, whatever their
original acceleration site. Even if all UHECR were generated at the
nucleus, we would not expect a source like Cen A to appear
`point-like' at the resolution of the PAO, so we cannot use the
observed large-scale excess of UHECR around Cen A to argue that
acceleration takes place either wholly or even partly in the giant
lobes. This limitation should be borne in mind in what follows.

The remainder of the paper is structured as follows. In Section
\ref{constraints} I show that the requirement that the lobes can
confine high-energy particles gives a potentially interesting
constraint on their radio luminosity, and argue that this means that
if the PAO UHECR are protons they are likely to originate in a small
number of bright nearby radio galaxies, all probably nearby
well-studied objects. In Section \ref{particle} I discuss our best
existing constraints on the particle content of FRI lobes and the
implications for cosmic ray acceleration and composition. Finally, in
Section \ref{outlook} I discuss the implications of a picture in which
particles are accelerated in radio galaxy lobes for the future of
cosmic ray astronomy. Throughout the paper I use a cosmology in which
$H_0 = 70$ km s$^{-1}$ Mpc$^{-1}$, $\Omega_{\rm m} = 0.3$ and
$\Omega_\Lambda = 0.7$. The distance to Cen A is taken to be 3.7 Mpc
(the mean of 5 distance estimates given in Ferrarese \etal\ 2007).

\section{Constraints on the radio emission}
\label{constraints}

In H09 we argued that stochastic acceleration by large-scale
  magnetic turbulence (as discussed by, e.g., Stawarz \& Petrosian
  2008) imposes a condition on the energy, acceleration region radius
  and magnetic field strength that is equivalent to the classical
  particle confinement condition. This comes about because the
  acceleration timescale, assuming Bohm diffusion, is $\sim 10r_{\rm
    L}/c$, where $r_{\rm L}$ is the Larmor radius, while the timescale
  for diffusive escape from the lobes is $\sim 3R^2/r_{\rm L}c$:
  equating these two gives $r_{\rm L} \sim R$ for the highest-energy
  cosmic rays that can be accelerated efficiently, which is simply the
  confinement condition. Ignoring numerical factors of order unity,
  therefore, we can consider the confinement condition as giving us
  the (best-case) estimate of the upper limit on the cosmic ray
  energy. In this section I demonstrate that the requirement that the
lobes be capable of confining UHECR at the energies observed gives
rise to an interesting constraint on the combination of the radio
luminosity and size of the lobes.

The confinement criterion for a particle of energy $E_p$ is that the
gyroradius $r_{\rm L}$ be less than the size of the region $R$; in
other words, in SI units,

\begin{equation}
R > {{\gamma_p m_0 c}\over{ZeB}} = {E_p\over{ZeBc}}
\label{confinement}
\end{equation}
where $Z$ is the nuclear charge and $e$ is the charge on the proton.
We consider a spherical radio lobe with radius $R$ and a uniform
magnetic field strength and electron energy density. Let the electron
energy distribution be given by $N(E_e)$ and let the magnetic field
be a factor $\epsilon$ away from equipartition, so that

\begin{equation}
U_e = \int_{E_{\rm min}}^{E_{\rm max}} E_e N(E_e) {\rm d}E_e = \epsilon U_B = \epsilon {B^2 \over{\mu_0}}
\label{edens}
\end{equation}
Here, as in H09, we are assuming `true' equipartition between the
electron energy spectrum and the magnetic field, by integrating over
all electron energies [following Myers \& Spangler (1985) and
  Hardcastle, Birkinshaw \& Worrall (1998)] rather than between the
energies corresponding to a pair of observed frequencies as in
`classical' equipartition. The differences between the two
equipartition formulae are discussed in more detail by Brunetti, Setti
\& Comastri (1997) and Beck \& Krause (2005). Beck \& Krause (2005)
show that the `classical' formula can lead to a significant
underestimate of the field strength, and thus to the ability to
confine high-energy particles, in radio galaxy lobes.

The simplest electron energy distribution we can consider is then a power
law in energy, i.e. $N(E_e) = N_0 E_e^{-s}$, with a minimum and
maximum energy given by $E_{\rm min}$ and $E_{\rm max}$ respectively.
This allows us to solve the integral of eq.\ \ref{edens} analytically.
Let

\begin{equation}
I = \int_{E_{\rm min}}^{E_{\rm max}} E_e E_e^{-s} {\rm d}E_e =
\left\{
\begin{array}{ll}
\ln(E_{\rm max}/E_{\rm min})&s=2\\
{1\over{2-s}} \left[E^{(2-s)}_{\rm max}-E^{(2-s)}_{\rm
    min}\right]&s\neq 2\\
\end{array}
\right .
\label{integ}
\end{equation}
The total energy in electrons is then $N_0 I$. However,
shock-acceleration models predict that the electron energy spectrum
should actually be described by a power law in momentum, i.e. $n(p) =
n_0 p^{-s}$ (e.g. Blandford \& Ostriker 1978). In this case, the
electron energy integral becomes
\begin{equation}
I = c^{1-s} m_e c^2 \int_{p_{\rm min}}^{p_{\rm max}} \left[ \left
  (1+{{p^2}\over{m_e^2c^2}}\right)^{1/2} - 1\right ] p^{-s} {\rm d}p
\label{mom-integ}
\end{equation}
(where $p_{\rm min}$ and $p_{\rm max}$ are the momenta corresponding
to the energies $E_{\rm min}$ and $E_{\rm max}$ respectively, and the
leading factor accounts for the difference between the normalizations
$N_0$ and $n_0$ in energy and momentum). It is most convenient to
evaluate the integral of equation \ref{mom-integ} numerically, though
clearly it converges to the analytical solutions of equation
\ref{integ} in the limit that $E_{\rm min} \gg m_e c^2$. We comment
below on the differences that arise when using these two values of
$I$.

\begin{figure}
\epsfxsize 8.5cm
\epsfbox{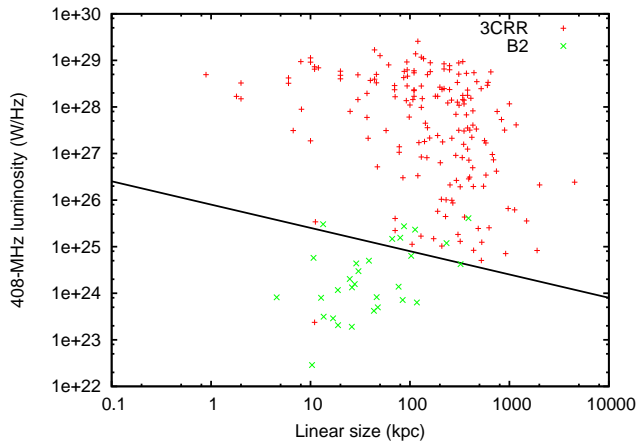}
\caption{The luminosity-size constraint of eq.\ \ref{llimit} for
  $\epsilon=1$, $E_{20}/Z = 1$ applied to both lobes (i.e. $D = 4R$,
  $L_T = 2L$) and plotted over
  the luminosity-size diagram for the B2 bright sample (LAS data from Fanti
  \etal\ 1987, flux densities, redshifts and spectral indices as
  tabulated by Hardcastle \etal\ 2003) and 3CRR sample (data from
    http://3crr.extragalactic.info/ ). Only sources above the solid
  line can confine UHECR with energies of $10^{20}$ eV; sources below the line
  cannot. While almost all the powerful 3CRR sources can, in principle, many
  of the lower-luminosity B2 sources cannot. Note that this figure is
  illustrative only; only a few of the lowest-luminosity objects in
  this plot are within the GZK cutoff and none are in the southern sky
  to which the PAO is most sensitive. No attempt has been made to take
  projection into account or to model the actual physical sizes of the
  lobes.}
\label{line}
\end{figure}

Now for a power-law electron energy distribution the volume emissivity
in synchrotron emission at a given (rest-frame) frequency $\nu$ may be written
\begin{equation}
J(\nu)  = C(s) N_0 \nu^{-{{(s-1)}\over 2}} B^{{(s+1)}\over 2}
\label{jnu}
\end{equation}
(Longair 1994 eq. 18.49) where
\begin{equation}
\begin{split}
C(s)&={{{\sqrt 3} e^3}\over{8 \pi \epsilon_0 c m_e (s+1)}}
\left({{2\pi m_e^3c^4}\over{3e}}\right)^{-(s-1)/2}\\
&\times{{\sqrt\pi \Gamma(\frac{s}{4} +
    \frac{19}{12}) \Gamma(\frac{s}{4} -
    \frac{1}{12}) \Gamma(\frac{s}{4} +
    \frac{5}{4})}\over{\Gamma(\frac{s}{4} + \frac{7}{4})}}\\
\end{split}
\end{equation}
for an isotropic pitch angle distribution. The frequency dependence in
eq.\ \ref{jnu} expresses the well-known relationship between $s$ and
the synchrotron spectral index $\alpha$. This result is valid both for
the truncated power-law distribution used in equation \ref{integ} and
for the power law in momentum described in equation \ref{mom-integ} so
long as the chosen observing frequency lies in a region where the
electron spectrum is a power law (i.e. not too close to the
high-energy or low-energy cutoff or to regions where $E \sim m_{\rm e}
c^2$). Since from equations \ref{edens} and \ref{integ} we have
\begin{equation}
\epsilon {B^2\over {2\mu_0}} = N_0 I
\label{bi}
\end{equation}
we can now use equation \ref{bi} to eliminate the electron spectral
normalization $N_0$ from equation \ref{jnu}:
\begin{equation}
\begin{split}
J(\nu)&=C(s) \epsilon {B^2\over {2I\mu_0}} \nu^{-{{(s-1)}\over 2}}
B^{{(s+1)}\over 2}\nonumber\\
&={C(s)\epsilon\over{2I\mu_0}} \nu^{-{{(s-1)}\over 2}}
B^{{(s+5)}\over 2}
\end{split}
\label{jnu2}
\end{equation}
Now rewriting eq. \ref{confinement} as $B > E_p/ZeRc$, we can
eliminate $B$ from equation \ref{jnu2}, turning it into an inequality:
\begin{equation}
J(\nu) > {C(s)\epsilon\over{2I\mu_0}} \nu^{-{{(s-1)}\over 2}}
\left({E_p \over {ZeRc}}\right)^{{(s+5)}\over 2}
\end{equation}
Finally, we can turn the emissivity $J(\nu)$ into an observable
quantity by noting that $L(\nu) = {4\over3}\pi R^3 J(\nu)$, so that
\begin{equation}
L(\nu) > {2\pi C(s)\epsilon\over{3I\mu_0}} \nu^{-{{(s-1)}\over 2}}
\left({E_p \over {Zec}}\right)^{{(s+5)}\over 2} R^{-{{(s-1)}\over 2}}
\label{l-ineq}
\end{equation}
We have derived a limit on the {\it luminosity} and size of a lobe
that is (marginally) capable of confining a particle of energy $E_p$
and charge $Z$. For conventional values of $s$ (in the range 2--3)
note the strong dependence of the luminosity on $E_P/Z$ (rigidity),
the linear dependence on the equipartition factor $\epsilon$, and the
relatively weak dependence on source size $R$, which is in the sense
that a lower luminosity is required for a larger size. For known
$\epsilon$, $s$ and $E_p/Z$, eq.
\ref{l-ineq} defines a line in the conventional radio luminosity/size
diagram for radio galaxies separating those that can accelerate such
particles from those that cannot (Fig.\ \ref{line}), if we adopt a
model such as that of H09 in which efficient stochastic particle
acceleration is possible in the lobes.

What are the implications for the population of radio galaxies that
can accelerate UHECR? First of all, we can substitute physical
constants into eq.\ \ref{l-ineq} to obtain a
relationship in useful units. If we take $s=2$, $E_{\rm min} = 5$ MeV,
$E_{\rm max} = 5$ GeV, $\nu=408$ MHz, $R =
r_{100}
\times 100$ kpc, $E = E_{20} \times 10^{20}$ eV, and use the
  numerically calculated expression for $I$ based on a power-law
  distribution in momentum, then we obtain
\begin{equation}
L_{408}>2.0 \times 10^{24} \epsilon \left(E_{20}\over Z\right)^{7/2}
r_{100}^{-1/2}\ {\rm W\ Hz^{-1}}
\label{llimit}
\end{equation}
What restrictions does this put on the population of radio galaxies
capable of accelerating UHECR? We can begin by turning this into a
strict limit on luminosity by imposing the observationally-based limit
that $R < 250$ kpc (i.e. $r_{100} < 2.5$) since we know that very few
radio galaxies exceed 1 Mpc in size. To compare to total radio
luminosity we must also scale up by a factor 2, since so far we have
only been considering the luminosity of a single lobe. This gives us a
strict lower limit on $L_{408}$ of $2.5 \times 10^{24}$ W Hz$^{-1}$
for $\epsilon=1.0$, $E_{20} = 1$, $Z=1$. Immediately we see
(Fig.\ \ref{line}) that only reasonably luminous radio galaxies can
accelerate UHECR to these energies; the Fanaroff-Riley break is at
$\sim 3 \times 10^{25}$ W Hz$^{-1}$ at 408 MHz. (Centaurus A, with a
408-MHz luminosity $\sim 3 \times 10^{24}$ W Hz$^{-1}$, just satisfies
this criterion, as we would expect given the results of H09.) If we
compare with a recent determination of the local radio luminosity
function, such as that by Mauch \& Sadler (2007), we find that within
a sphere of radius 100 Mpc (the right order of magnitude for $10^{20}$
eV protons) we expect $\sim 20$ radio galaxies satisfying the
luminosity criterion alone, of which not all will satisfy the size
criterion (and of course only roughly half of which will be visible to
the PAO). Thus for the parameters we have used here we see that radio
galaxies capable of accelerating protons to $10^{20}$ eV will be rare.
In addition, since their luminosities are large and their distances
constrained, their fluxes are known ($>2$ Jy at 408 MHz) and so we can
say that all such objects are probably bright, well-studied local
radio galaxies.

Let us now consider varying some of the assumptions in the
calculations above.

\subsection{Power-law index and minimum energy}

\begin{figure}
\epsfxsize 8.5cm
\epsfbox{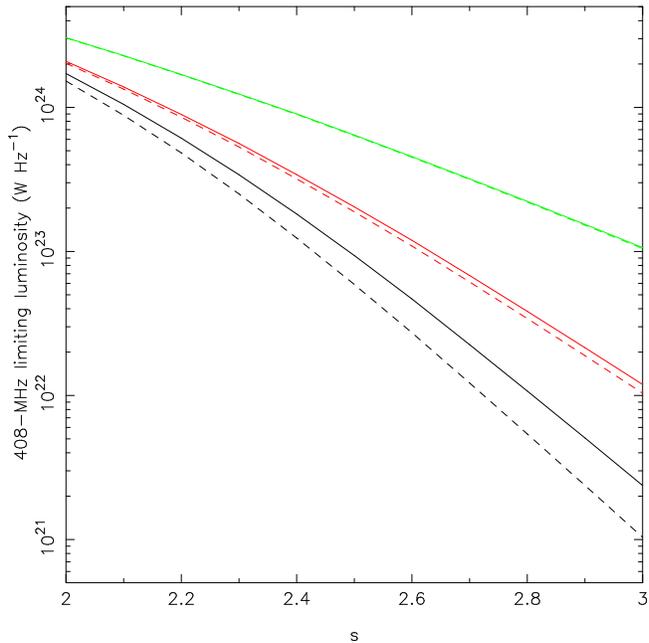}
\caption{The normalizing luminosity of equation \ref{llimit} as a
  function of the power-law index $s$ of the energy/momentum power
  law. The lower (black) curve shows $E_{\rm min} = 0.5$ MeV, the
  middle (red) curve shows $E_{\rm min} = 5$ MeV, and the upper
  (green) curve shows $E_{\rm min} = 50$ MeV. The solid lines show the
  luminosity for an assumed power-law distribution in momentum and the
  dashed lines show a power law in energy. The clearest trend is a
  decrease in the limiting luminosity with increasing $s$. We also see
  that the effect of changing $E_{\rm min}$ is very limited for $s=2$
  but very significant by the time $s=3$. The effect of incorrectly
  assuming a power-law distribution in energy is only significant for
  the lowest value of $E_{\rm min}$.}
\label{splot}
\end{figure}

The normalizing luminosity in equation \ref{llimit} has a relatively
strong dependence on the power-law index $s$. This reflects the fact
that the electron energy density, and thus the magnetic field
strength, is dominated by the low-energy electrons, while it is
high-energy electrons that produce the observed radio emission at our
normalizing frequency. There is also a dependence on the minimum
energy $E_{\rm min}$ which is stronger for larger $s$. These
dependencies are illustrated in Fig. \ref{splot}. Values of $s$ close
to 2.0 are predicted in shock acceleration models and appear to be
consistent with observation (cf.\ Young \etal\ 2005). Fig. \ref{splot}
shows that the normalizing luminosity is not greatly affected for
values close to 2, say $s=2.2$, so that the number of potential
accelerating sources is probably not greatly affected by our
uncertainties on this parameter or on the appropriate value of $E_{\rm
  min}$.

The dependence on $E_{\rm max}$ is always weak, and so it is
unnecessary to put in a more realistic electron energy distribution
with spectral steepening below $E_{\rm max}$.

\subsection{Particle energy}

The number of radio galaxies capable of accelerating UHECR protons to
high energies is a very strong function of $E_{20}$. Even for $E_{20}
= 2$, the expected number of radio galaxies in the southern sky within
100 Mpc that satisfy eq.\ \ref{llimit}, neglecting the size
constraint, is less than 1 (and here we also neglect the steep
decrease in the appropriate radius to use due to the energy-dependent
GZK cutoff). By contrast, if we set $E_{20} = 0.6$, there are perhaps
40 radio galaxies in the southern sky that are in principle capable of
accelerating protons to those energies, again neglecting the size
constraint. Effectively, therefore, this model for UHECR acceleration
predicts a very steep cutoff in the integrated {\it source} spectrum of UHECR
protons which, by chance, occurs at energies close to the energy at
which GZK effects become significant, and which therefore reinforces
the effect of the GZK cutoff.

\subsection{Equipartition}

In the calculations above I have used $\epsilon = 1$, corresponding to
equipartition between radiating particles and magnetic field, which is
consistent with the known constraints from inverse-Compton radiation
from Cen A. If there were an energetically dominant population of
non-radiating charged particles, such as protons, we would expect the
magnetic field to be in equipartition with those and so $\epsilon \ll
1$. On the other hand, the evidence in those FRII sources in which
inverse-Compton modelling has been possible is that the magnetic field
strength is typically somewhat below the equipartition value, implying
$\epsilon > 1$. Values of $\epsilon \gg 1$, implying very low
$B$-field strengths for a given observed synchrotron luminosity,
obviously make it very hard for lobes to accelerate UHECR. Values $\ll
1$ make it easier, but given the rather flat radio luminosity function
do not immediately fill the sky with UHECR-emitting radio galaxies:
for example, substituting $\epsilon = 0.1$, $E_{20} = 1$, $Z=1$,
$r_{100}=2.5$ in eq.\ \ref{llimit} and integrating over the luminosity
function gives around 50 radio galaxies in the southern sky that meet
the luminosity constraint.

\subsection{Relativistic turbulence}

The acceleration timescale and therefore the efficiency of UHECR
acceleration in the acceleration model of H09 depend on the presence
of strong magnetic turbulence in the lobes: we require $U_{\rm turb}
\sim U_0$ where $U_{\rm turb}$ and $U_0$ are the energy densities in
the turbulent magnetic field component and the unperturbed component
respectively. Once $U_{\rm turb}$ becomes much less than $U_0$
turbulent acceleration will be very much less efficient. Magnetic
turbulence is presumably generated by large-scale hydrodynamic
processes and therefore relies on a continued energy supply by the
jet. In lobes where the jet is disconnected the turbulence will decay
on timescales which may be as short as $R/c$. Since the energy density
in relativistic turbulence may therefore depend on local details of
the coupling between the jets and the large-scale lobes, it is clearly
therefore possible to imagine a situation in which the particle
acceleration efficiency varies from source to source, depending on
such factors as the large-scale morphology of the lobes and the
presence or absence of strong jet-lobe interactions (such as jet
termination shocks). We necessarily cannot take account of this in the
simple models presented here, but it should be borne in mind that the
luminosity/size cutoffs apply only in sources in which relativistic
turbulence can efficiently be maintained; in particular, disconnected
lobes (possibly even including the S giant lobe of Cen A; see H09) are
likely to be unable to accelerate UHECR.

\subsection{Composition}
\label{heavies}
As with all UHECR acceleration models, it is much easier to accelerate
nuclei, with $Z>1$, than protons to a given energy. If we naively
substitute $\epsilon = 1.0$, $E_{20} = 1$, $Z=26$ (iron),
$r_{100}=2.5$ into eq.\ \ref{llimit}, then the luminosity limit comes
down by nearly 5 orders of magnitude and we find that practically
every radio galaxy in the sky is a potential UHECR source, although in
practice the size constraint will still impose some limitations. A
self-consistent model for heavy nucleus acceleration in the lobes
would need to take account of losses to photodisintegration within the
lobes themselves --- both the acceleration and loss timescales are
shorter for nuclei than for protons --- but such a model is beyond the
scope of this paper. Here we have simply to note that the strong
rigidity dependence of eq.\ \ref{l-ineq} means that the composition of
the baryons available to accelerate in the source will have a strong
effect on the predicted composition, energy spectrum and arrival
positions of UHECR in this model (all of course modified by
propagation effects; e.g. Hooper \& Taylor 2009). We discuss the
available constraints on composition in the next section.

\section{Particle content}
\label{particle}

The preceding sections have shown a strong dependence of the
predictions of a model in which UHECR are accelerated in radio galaxy
lobes on the source composition of the particles in the lobes, and so
at this point it is appropriate to comment on the known
constraints on the particle content of lobes, and to ask what sources
of (1) protons and (2) heavy nuclei are available on these scales.

We do not know whether jets in radio-loud AGN are electron-proton or
electron-positron in their initial composition. There is some evidence
in FRII radio galaxies that the lobes are not {\it dominated}
energetically by protons --- see Croston \etal\ (2005) for the
argument --- but there is no way in these FRII lobes to rule out the
possibility that a relativistic proton population has energy roughly
comparable to that in the electrons and magnetic field, whose energy
densities can be measured. A fortiori, we do not know the expected
fraction of heavy nuclei in these lobes, since even if the jets
contain protons, it is not clear at what point in the jet generation
process they get there.

The situation in FRI radio galaxies has been known for many years to
be more interesting. Here inverse-Compton measurements are not in
general available, but the minimum pressures (approximately equivalent
to the assumption of equipartition between field and electrons alone)
in the large-scale lobes or plumes can be several orders of magnitude
below those of the external medium (see Hardcastle \etal\ 2007b and
references therein) and, since inverse-Compton constraints rule out
electron dominance by very large factors and energetic dominance by
magnetic field seems a priori implausible, it is conventional to
suggest that the missing pressure is supplied by a population of
non-radiating particles (perhaps with the magnetic field in
approximate equipartition, implying $\epsilon \ll 1$). We have
recently argued (Croston \etal\ 2008; Croston \& Hardcastle, in prep.)
that there is some evidence that these particles are the same particle
population that is required to be entrained to decelerate the
kpc-scale jet; this requires a means of efficiently heating or
accelerating these particles to make them provide the required
pressure. Entrainment is interesting here because it does give us some
constraints on the expected abundance of the heavy particle
population, which should be similar to that of the external medium
(i.e. roughly 1 iron nucleus per $10^5$ protons for an assumed 0.3
solar abundance). It is important to note, though, that if the
entrained particles were thermal and dominated the energetics, then
the Alfv\'en speeds in the lobes would be $\ll c$ and the efficiency
of stochastic acceleration in the lobes would be greatly reduced
(H09; O'Sullivan \etal\ 2009). To avoid this, we would need the entrained
particles to be relativistic and to participate in the equipartition
process so that the energy densities in magnetic field and baryons
were comparable. There is as yet no direct evidence that rules this
picture out (see Croston \& Hardcastle, in prep., for more discussion
of constraints on the state of the entrained material). In addition,
we have shown that the necessity for entrainment varies from source to
source, even among FRIs (Croston \etal\ 2008). It is certainly
possible that jets in the low-power sources are initially
electron-positron and that the amount of material entrained, at least
in some sources, is enough to provide the seed population for
stochastic acceleration of baryons in the lobes while not being so
much that stochastic acceleration is inefficient.

\section{Predictions and the outlook for cosmic ray astronomy}
\label{outlook}

One of the attractive features of the model proposed in H09 and
discussed in this paper is that, at least
superfically, it makes some simple testable predictions. If the UHECR
mapped by the PAO are protons, which would be implied by a genuine
detection of correlation on the sky with the positions of distant
objects (perhaps excluding Cen A; see below) then I have shown above
that the lobe acceleration model would imply that their local sources are
physically large ($>100$ kpc), luminous, relatively rare radio
galaxies. A couple of hints that this is so are already seen, firstly in the
apparent excess of events around Cen A, and secondly in the work of Nagar
\& Matulich (2008) who found a correlation between the positions of
extended radio galaxies and the arrival positions of the PAO UHECR. 
With the eventual release of updated PAO positional data it should be
possible to make a systematic investigation of all possible radio-loud
sources of UHECR.

However, the situation is complicated by the current uncertainty about
the composition of the PAO cosmic rays. Current measurements of the
mean and RMS depth of shower maximum imply a large fraction of heavy
or intermediate mass nuclei, and certainly do not appear consistent
with a pure-proton spectrum (Abraham \etal\ 2010), although it is
important to note that the HiRes results are quite different (e.g.
Aloisio, Berezinsky \& Gazizov 2009). If we are required to accelerate
nuclei with $Z>1$, then the predictions change in two crucial ways:
firstly, the number of potential radio-galaxy UHECR sources increases
rapidly with increasing $Z$, as discussed above (Section
\ref{heavies}); secondly, it becomes increasingly unlikely that a
spatial correlation will be observed between the arrival directions of
the UHECR and the positions of their sources, due to the larger
deflection of $Z>1$ UHECR in intergalactic magnetic fields. At this
point a model that intended to reproduce the observations would need
to take account of (1) the spatial distribution of the radio galaxy
sources throughout the GZK volume and perhaps beyond; (2) the
distribution of the physical conditions in their lobes; (3) the
intrinsic UHECR energy and composition spectrum; (4) propagation
losses for the various species of UHECR; and (5) deflection in the
Galactic and intergalactic magnetic fields. The tools to do (1) and
(2) are available and to some extent presented in this paper; (3)
remains very uncertain, though we have some constraints (see Section
\ref{particle} above); and (4) and (5) are in principle possible
(e.g., Hooper \& Taylor 2009), although crucial elements remain
uncertain. Putting all five elements together would be a major effort
which may not yet be justified by the state of the data, but detailed
modelling like this will be the way forward if we are to start doing
serious astrophysics with UHECR observations.

Having said this, some limiting cases of the model if the PAO UHECR
are heavy or intermediate-mass nuclei are relatively easy to imagine.
Aloisio \etal\ (2009) discuss what they call the `disappointing model'
for the PAO results in which rigidity-dependent acceleration (as is
implicit in the lobe-acceleration picture) together with an
acceleration cutoff for protons at $\sim 10^{19}$ eV leads to a
steadily increasing fraction of heavy nuclei with increasing energy
across the PAO band, as observed. Given that we require the most
favourable assumptions to make radio galaxies accelerate protons up to
$10^{20}$ eV, it is easy to imagine that the high-energy cutoff might
be reduced by an order of magnitude or so (e.g., by reducing the
  maximum energy placed into large-scale magnetic turbulence), so
that they might provide the `disappointing' population required by
Aloisio \etal The detection of an enhanced count rate around Cen A
would then be explained by its proximity, which has the effect that
few-$Z$ particles are deflected by the Galactic and intergalactic
magnetic fields by only a few (up to 10) degrees. However, the UHECR
from all other sources would be scattered by much larger angles and it
would never be possible to identify them with their parent radio
galaxies. While this model would be slightly less disappointing than
the limiting case suggested by Aloisio \etal , Cen A would remain the
only detectable UHECR source in the sky.

\section{Summary}

The principal results of this paper may be summarized as follows:

\begin{enumerate}
\item Stochastic acceleration of UHECR in the large-scale lobes of
  radio galaxies may be possible, but there are strong (though
  model-dependent) constraints on the properties of the radio galaxies
  that can accelerate them to the highest energies ($10^{20}$ eV).
\item These constraints imply that only a small number of local radio
  galaxies can be involved in the acceleration of UHECR, if the UHECR
  are protons, and that UHECR energies will cut off steeply around the
  energies currently being observed by the PAO; this model is testable
  in principle using existing radio surveys and up-to-date UHECR
  arrival positions, and is consistent with much of the available
  data.
\item However, if UHECR are heavy nuclei with $Z > 1$, as suggested by
  the latest PAO composition results, then many more radio
  galaxies can be sites of UHECR acceleration, and it may be that the
  nearest radio galaxy, Cen A, will be the only identifiable source in
  the cosmic-ray sky.
\end{enumerate}

\section*{Acknowledgements}
This work was first presented at the Trondheim workshop on `Searching
for the origins of cosmic rays' in June 2009, and I am indebted to the
organizers for inviting me and to many participants there for helpful
comments. I also owe a debt of gratitude to \L ukasz Stawarz and Teddy
Cheung, without whom the discussion of cosmic ray acceleration in H09
would have been extremely limited. The paper was substantially
  improved as a result of comments from an anonymous referee. I
acknowledge generous financial support from the Royal Society through
the University Research Fellowships scheme.

\end{document}